\documentclass[twocolumn,superscriptaddress,showpacs,preprintnumbers,amsmath,amssymb,pre]{revtex4}
\usepackage{amsmath}
\usepackage{amssymb}
\usepackage{graphicx}
\usepackage{dcolumn}% Align table columns on decimal point
\usepackage{subfigure}
\usepackage[bookmarks=true,colorlinks=false,urlcolor=blue]{hyperref} 
\usepackage{wrapfig}
\usepackage{sidecap}
\sidecaptionvpos{figure}{c}
\usepackage{bm}% bold math
\usepackage{float}

\begin{document}

\title{Random close packing in protein cores} 

\author{Jennifer C. Gaines} 
%\email{jennifer.gaines@yale.edu}
\affiliation{Program in Computational Biology and Bioinformatics, Yale University, New Haven, Connecticut, 06520}
\affiliation{Integrated Graduate Program in Physical and Engineering Biology (IGPPEB), Yale University, New Haven, Connecticut, 06520}
\author{W. Wendell Smith }
\affiliation{Department of Physics, Yale University, New Haven, Connecticut, 06520}
\author{Lynne Regan}
\affiliation{Program in Computational Biology and Bioinformatics, Yale University, New Haven, Connecticut, 06520}
\affiliation{Integrated Graduate Program in Physical and Engineering Biology (IGPPEB), Yale University, New Haven, Connecticut, 06520}
\affiliation{Department of Molecular Biophysics \& Biochemistry, Yale University, New Haven, Connecticut, 06520}
\affiliation{Department of Chemistry, Yale University, New Haven, Connecticut, 06520}
\author{Corey S. O'Hern}
\affiliation{Program in Computational Biology and Bioinformatics, Yale University, New Haven, Connecticut, 06520}
\affiliation{Integrated Graduate Program in Physical and Engineering Biology (IGPPEB), Yale University, New Haven, Connecticut, 06520}
\affiliation{Department of Physics, Yale University, New Haven, Connecticut, 06520}
\affiliation{Department of Mechanical Engineering \& Materials Science,
Yale University, New Haven, Connecticut, 06520}
\affiliation{Department of Applied Physics,
Yale University, New Haven, Connecticut, 06520}

\begin{abstract}
Shortly after the determination of the first protein x-ray crystal
structures, researchers analyzed their cores and reported packing
fractions $\phi \approx 0.75$, a value that is similar to close
packing equal-sized spheres. A limitation of these analyses was the
use of `extended atom' models, rather than the more physically
accurate `explicit hydrogen' model. The validity of using the explicit
hydrogen model is proved by its ability to predict the side chain
dihedral angle distributions observed in proteins. We employ the
explicit hydrogen model to calculate the packing fraction of the cores
of over $200$ high resolution protein structures. We find that these
protein cores have $\phi \approx 0.55$, which is comparable to random
close-packing of non-spherical particles.  This result provides a
deeper understanding of the physical basis of protein structure that
will enable predictions of the effects of amino acid mutations and
design of new functional proteins.
\end{abstract}

\pacs{87.15.A-, 87.14.E-, 87.15.B-}

\maketitle

\begin{figure}[!htp]
\centering
\includegraphics[width= 3.25in]{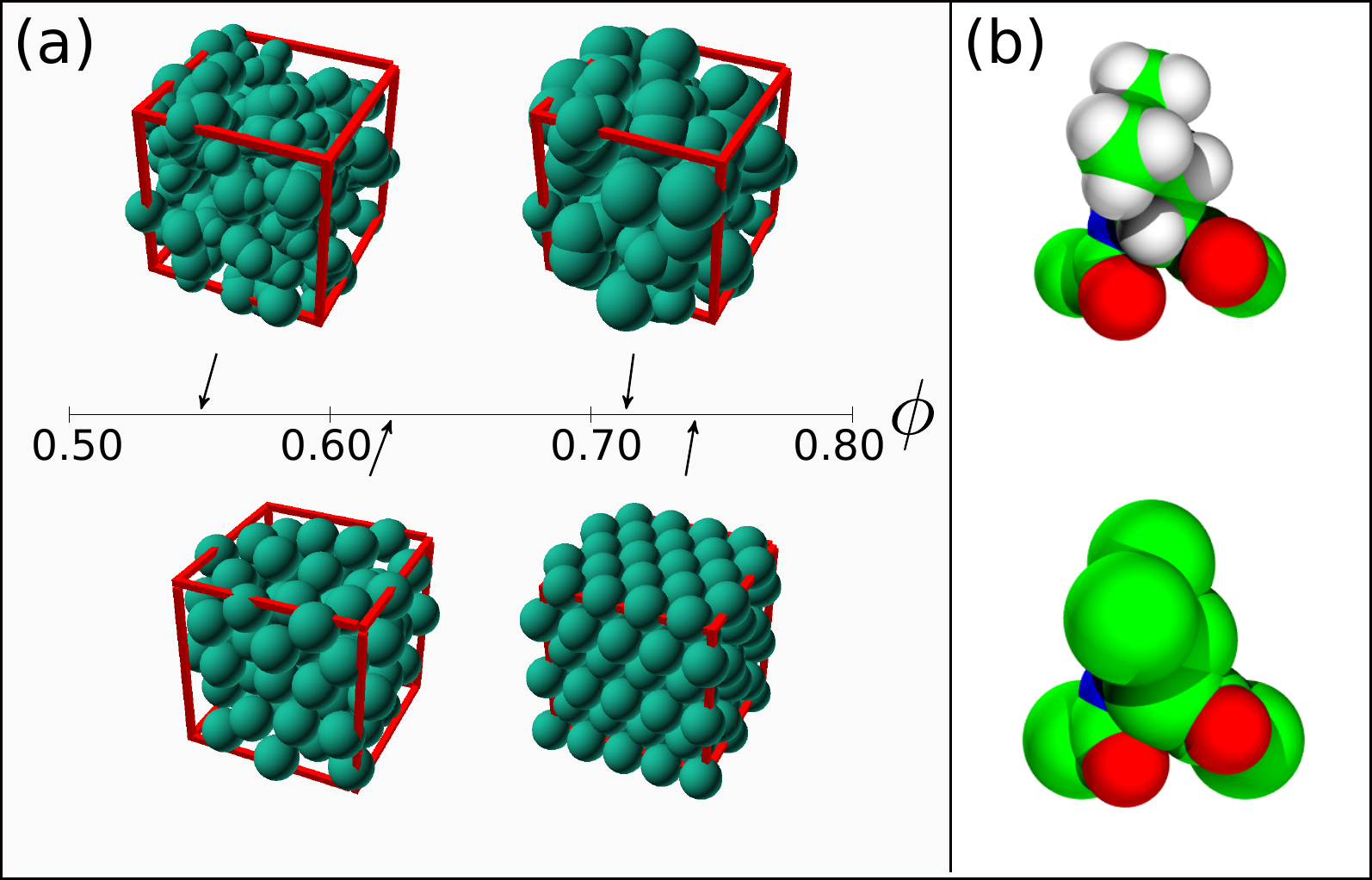}
\caption{(a) Visualization of core residues for a typical protein (Carboxyl Proteinase) in
the Dunbrack database of crystal structures
using explicit hydrogen (top left, $\phi \approx 0.55$) and extended
atom (top right, $\phi \approx 0.72$) models compared to random
close (bottom left, $\phi_{\rm RCP} \approx 0.64$) and face centered
cubic packed (bottom right, $\phi_{\rm FCC} \approx 0.74$) systems
with equal-sized spheres. (b) Leu residue with each atom
represented as a sphere using the explicit hydrogen (top) 
and extended atom (bottom) models. The atom types are shaded 
green (carbon), red (oxygen), blue (nitrogen), and gray (hydrogen).}
\label{fig:phi_line1}
\end{figure}

It is generally accepted that hydrophobic cores of proteins are
tightly packed. In fact, many biology textbooks state that the packing
fraction of protein cores is similar to that of densely packed
equal-sized spheres with $\phi = 0.74$~\cite{kyte}.  Using a more
accurate stereochemical representation, we show that the packing
fraction of protein hydrophobic cores is $\phi \approx 0.55$
(Fig. \ref{fig:phi_line1} (a) top left), which is similar to values
for random close packing of non-spherical particles~\cite{nolan,jia},
not close packing of equal-sized spheres (Fig. \ref{fig:phi_line1} (a)
bottom right).

\begin{figure*}
\centering
\includegraphics[width = 7in]{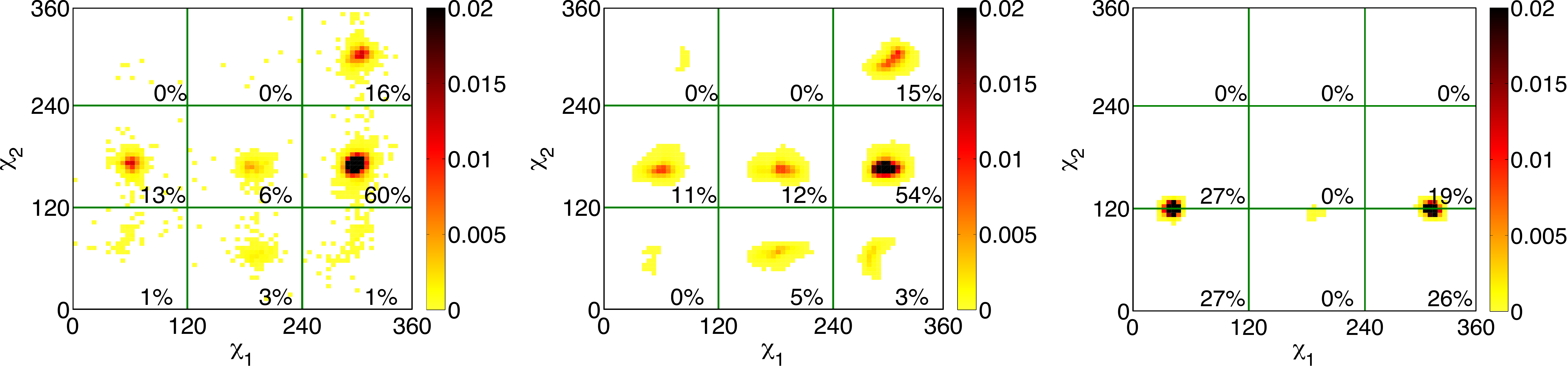}
\caption{(left) The observed side chain dihedral angle probability
distribution $P(\chi_1,\chi_2)$ for Ile residues in the Dunbrack
database of protein crystal structures. We also show $P(\chi_1,\chi_2)$ 
predicted by the hard-sphere dipeptide mimetic
model for Ile using the (center) explicit hydrogen and (right) extended atom
representations. For the extended atom model, we used the
atomic radii in the original work by Richards~\cite{Richards}. The 
probabilities increase from light to dark. The percentages give 
the fractional probabilities that occur in each of the nine 
square bins.}
\label{hs_dist}
\end{figure*}

The most influential study of packing in protein cores was performed
by Richards in 1974~\cite{Richards}. He used Voronoi tessellation to
calculate the packing fraction in the hydrophobic cores of two of the
few proteins whose crystal structures had been determined at that time
- lysozyme and ribonuclease S. He reported that the mean packing
fraction of the two protein cores is $\phi_0 \approx 0.75$.  More
recent studies have obtained similar values for the packing fraction
using larger data sets of protein
cores~\cite{Dill,Fleming,Rother,Ponder}. We believe that the reason
these prior studies have calculated such high values for the packing
fraction of protein cores is that they use an `extended atom'
representation of the heavy atoms. In this representation, hydrogen
atoms are not included explicitly, rather the atomic radius of each
heavy atom is increased by an amount proportional to the number of
hydrogens that are bonded to it.  An extended atom representation is
often employed in computational studies of proteins because it
significantly decreases the calculational complexity.  In
Fig.~\ref{fig:phi_line1} (b), we compare the extended atom
representation of a Leu residue to one that includes hydrogen atoms
explicitly. It is clear that the extended atom and explicit hydrogen
representations of Leu possess different sizes and shapes.

In a 1987 paper on protein core re-packing, Ponder and
Richards~\cite{Ponder} stated that ``...the use of extended atoms was
not satisfactory. In order for the packing criteria to be used
effectively, hydrogen atoms had to be explicitly included..."  Ponder
and Richards argued that the extended atom model did not provide a
sufficiently accurate representation of the stereochemistry of amino
acids. In this manuscript, we examine the packing fraction of the
hydrophobic cores of a large number of proteins using the explicit
hydrogen representation, as Ponder and Richards ~\cite{Ponder}  and
others~\cite{reduce} advocate. We find that the average packing
fraction of protein cores is $\phi \approx 0.55$. We obtain similar
results from hard-sphere models of mixtures of residues that are
isotropically compressed to jamming onset. Knowing the correct packing
fraction of protein cores is important because one needs to know the
naturally occurring value to assess the effects of amino acid mutations, or to
design new proteins. Strong support for the validity of the explicit
hydrogen representation is that this model is able to reproduce the
observed side chain dihedral angle distributions of residues in
protein cores, whereas the extended atom representation does not.

To calculate the packing fraction of protein cores, we use the
`Dunbrack database' of high resolution protein crystal structures,
which is composed of $221$ proteins with resolution $\leq 1.0$ {\AA}, side
chain B-factors per residue $\leq 30$ {\AA}$^2$, and R-factor $\leq
0.2$~\cite{Dun_1a, Dun_1b}. In prior studies, we showed that
hard-sphere models of dipeptide mimetics with explicit hydrogens can
recapitulate the side chain dihedral angle distributions observed in
protein crystal structures~\cite{HS_Z2013, HS_Z2014, HS_Z2011,
  HS_Z2012, HS_C2014}.

For the hard-sphere model, each atom $i$ in a dipeptide mimetic is treated
as a sphere that interacts pairwise with all other non-bonded atoms
$j$ via
\begin{equation}
\label{vrlj}
U_{\rm RLJ}(r_{ij}) = \frac{\epsilon}{72}\left[1-\left(\frac{\sigma_{ij}}{r_{ij}}\right)^6\right]^2 \Theta(\sigma_{ij} - r_{ij}), 
\end{equation}
where $r_{ij}$ is the center-to-center separation between atoms $i$
and $j$, $\Theta(\sigma_{ij}-r_{ij})$ is the Heaviside step function,
$\epsilon$ is the energy scale of the repulsive interactions,
$\sigma_{ij} = (\sigma_i+\sigma_j)/2$, and $\sigma_i/2$ is the radius
of atom $i$.  A dipeptide mimetic is a single amino acid plus the
$C_{\alpha}$, $C$, and $O$ of the prior amino acid and the $N$, $H$,
and $C_{\alpha}$ of the next amino acid. Bond lengths and angles are
set to those in the Dunbrack database. Hydrogen atoms were added using
the REDUCE software program \cite{reduce}, which sets the bond lengths
for $C$-$H$, $N$-$H$, and $S$-$H$ to $1.1$, $1.0$ and $1.3$ {\AA},
respectively, and the bond angles to $109.5^\circ$ and $120^\circ$ for
angles involving $C_{\mathrm{sp^2}}$ and $C_{\mathrm{sp^3}}$ atoms.  Additional
dihedral angle degrees of freedom involving hydrogens are chosen to
minimize steric clashes~\cite{reduce}.

Predictions for the side chain dihedral angle distributions of a given
dipeptide mimetic are obtained by rotating each of the side chain dihedral angles
$\chi_1,\ldots,\chi_n$ and evaluating the total potential energy
$U(\chi_1,\ldots,\chi_n)=\sum_{i<j} U_{\rm RLJ}(r_{ij})$ and Boltzmann
weight
\begin{equation}
\label{probability}
P(\chi_1,\ldots,\chi_n) \propto e^{-U(\chi_1,\ldots,\chi_n)/k_BT}.
\end{equation}
We then average the Boltzmann weight over all dipeptide mimetic and normalize such that $\int
P(\chi_1,\ldots,\chi_n)d\chi_1,\ldots,d\chi_n=1$. We set the
temperature $k_BT <10^{-2}$ to be sufficiently small that we are in the
hard-sphere limit and $P(\chi_1,\ldots,\chi_n)$ no longer depends on
temperature.  The values for the six atomic radii ($C_{sp^3}$, $C_{\rm aromatic}$: $1.5$
{\AA}; $C_O$: $1.3$ {\AA}; $O$: $1.4$ {\AA}; $N$: $1.3$ {\AA}; $H$:
$1.10$ {\AA}; and $S$: $1.75$ {\AA}) were obtained by minimizing the
difference between the side chain dihedral angle distributions
predicted by the hard-sphere dipeptide mimetic model and those observed in protein
crystal structures for a small subset of amino acid types.  The atomic
radii are similar to values of van der Waals radii reported in earlier
studies~\cite{HS_Z2012,Rama,Bondi,CDC,Seeliger,Pauling,Porter,Chothia}. (See Supplemental Material.)

The packing fraction of each residue was calculated using
\begin{equation}
\label{phi_eq}
\phi = \frac{\sum \mathrm{V}_{i}}{\sum \mathrm{V}^v_{i}},
\end{equation}
where $\mathrm{V}_{i}$ is the `non-overlapping' volume of atom {\it
  i}, $\mathrm{V}^v_{i}$ is the Voronoi volume of atom {\it i}, and
the summation is over all atoms of a particular residue.  The
non-overlapping volume of each atom is obtained by dividing
overlapping atoms $i$ and $k$ by the plane of intersection between the two spheres. $\mathrm{V}^v_{i}$ for each atom was found
using a variation of the Voro++ software library~\cite{Vor}. Voronoi
cells were obtained for each atom using Laguerre tessellation, where
the placement of the Voronoi cell walls is based on the relative radii
of neighboring atoms (which is the same as the location of the plane 
that separates overlapping atoms).

We define core residues as those that are neither on the protein
surface nor on the surface of an interior void. We identify surface
and void atoms as those with empty space next to them. Points were
found that were greater than $1.4$ {\AA} (approximately the radius of a water molecule)
from the surface of all atoms in the protein using Monte Carlo
sampling. The closest atom to each of these points was designated as a
surface atom. For a residue to be considered a core residue, it must
not contain any surface atoms.  According to this definition and using
the explicit hydrogen representation, proteins in the Dunbrack
database had an average of $15$ core residues. Ala, Cys, Gly, Ile,
Leu, Met, Phe, and Val residues make up over $80\%$ of the protein cores.
However in our calculations of the packing fraction of protein crystal structures
we included all amino acid types.

\begin{figure}[!htp]
\centering
 \includegraphics[width=3in]{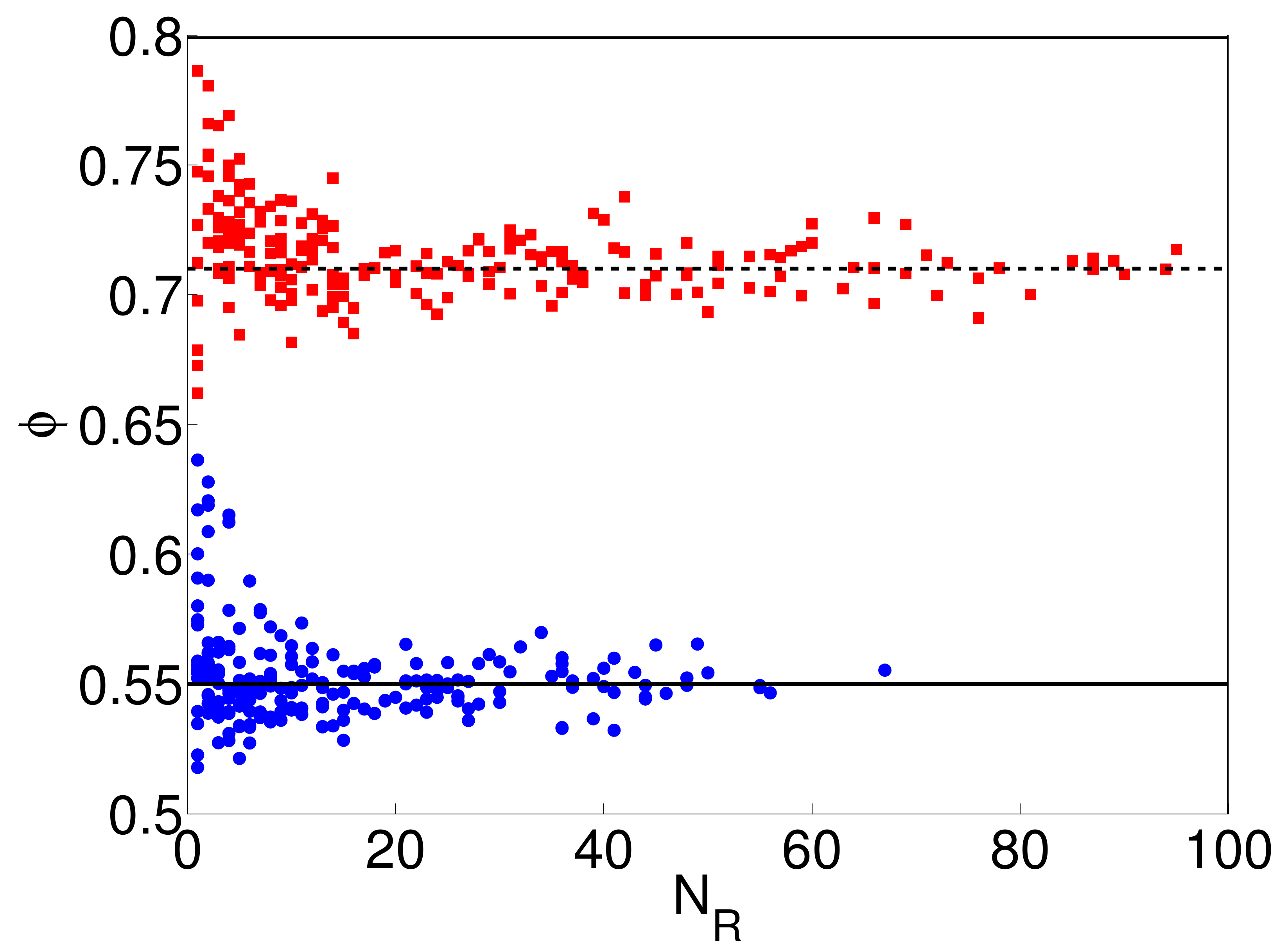}
\caption{(color online) A comparison of the packing fraction $\phi$
of the cores of proteins in the Dunbrack database as a function of
the number of core residues $N_R$ using the explicit hydrogen (blue
circles) and extended atom (red squares) representations.  More
residues are designated as core using the extend atom model ($25$ on
average) than using the explicit hydrogen model ($15$ on average).
The solid and dashed horizontal lines indicate $\langle \phi
\rangle_{EH} \approx 0.55$ and $\langle \phi \rangle_{EA} \approx
0.71$.}
\label{phi_length}
\end{figure}

We also performed similar packing analyses using the extended atom
representation with the same atom types and radii used by Richards
($N$: $1.7$ {\AA}, $O$: $1.4$ {\AA}, $O(H)$: $1.6$ {\AA}, $C$: $2.0$
{\AA}, and $S$: $1.8$ {\AA}) with the exception of C for the ring systems (Phe,
Tyr, Trp, Arg, and His) which was set to $1.7$ {\AA} ~\cite{Richards}. For both explicit hydrogen and extended atom
representations, we calculated $\phi$ for the core of a given
protein using Eq.~\ref{phi_eq} with the summation over all atoms of all
residues in the core. We also calculated the packing fraction for each residue
in the core with the summation over all atoms
in the residue.

In Fig.~\ref{hs_dist}, we compare the observed side chain dihedral
angle distributions for Ile residues in the Dunbrack database and the
predicted distributions from the hard-sphere dipeptide mimetic model using the
explicit hydrogen and extended atom representations. The observed
distribution for Ile (Fig.~\ref{hs_dist} (left)) possesses one strong
peak at $\chi_1=300^{\circ}$, $\chi_2=180^{\circ}$ and three minor
peaks at $\chi_1=300^{\circ}$, $\chi_2=300^{\circ}$,
$\chi_1=60^{\circ}$, $\chi_2=180^{\circ}$, and $\chi_1=180^{\circ}$,
$\chi_2=180^{\circ}$.  The side chain dihedral angle distribution for
Ile predicted using the hard-sphere dipeptide mimetic model with the explicit
hydrogen representation reproduces each of these features
(Fig.~\ref{hs_dist} (center)).  In contrast, the high probability
regions of $\chi_1$-$\chi_2$ space for the extended atom
representation of the Ile dipeptide mimetic occur near $\chi_1=60^{\circ}$,
$\chi_2=120^{\circ}$ and $\chi_1=300^{\circ}$, $\chi_2=120^{\circ}$,
which have extremely low probability in the observed distributions.
These results (and those shown in prior work for Val, Leu, Phe, Tyr, Thr, Ser, and Cys \cite{HS_Z2014}) show that the extended atom model of a dipeptide mimetic does
not reproduce the observed dihedral angle distribution, whereas the
explicit hydrogen model of a dipeptide mimetic does.

The results for the packing fraction analyses on core residues in all
proteins in the Dunbrack database are shown in Fig.~\ref{phi_length}.
For the explicit hydrogen representation, we find that the average
packing fraction in protein cores is $\langle \phi \rangle_{EH}
\approx 0.55 \pm 0.02$ (blue circles), with fluctuations that are
larger in proteins with small cores. This value is significantly lower
than that obtained using the extended atom representation, $\langle
\phi \rangle_{\rm EA} \approx 0.71 \pm 0.05 $ (red squares), which is
similar to $\phi_0 \approx 0.75$ reported in
Ref.~\cite{Richards}. (The slight difference between $\langle \phi
\rangle_{\rm EA}$ and $\phi_0$ is due to the higher resolution of the
Dunbrack database and that Richards averaged the local atomic packing
fractions rather than taking the ratio of the total volumes as in
Eq.~\ref{phi_eq}.)

\begin{figure}[!htp]
\centering
\includegraphics[width=3in]{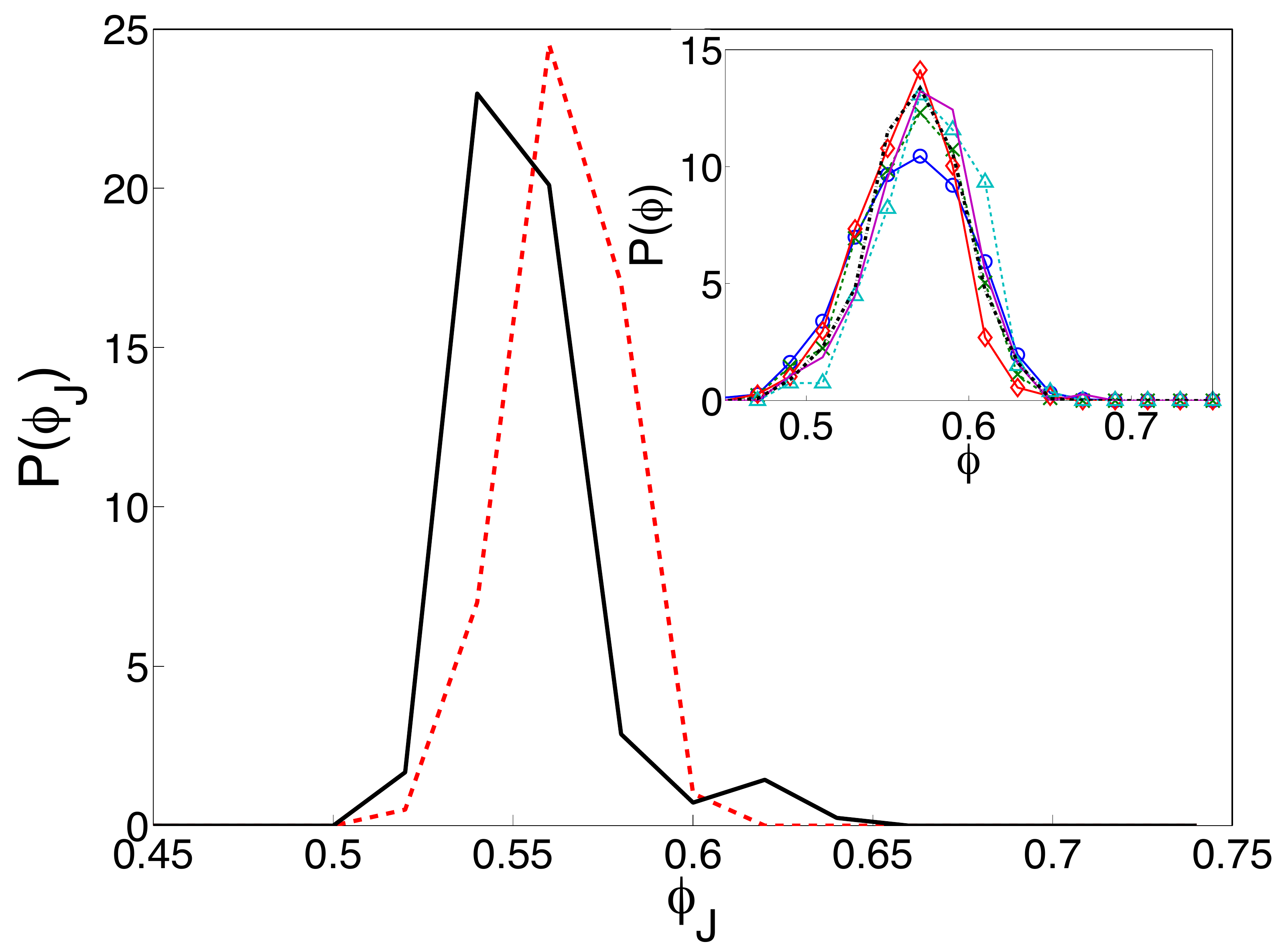}
\caption{The distribution of packing fractions $P(\phi_J)$ from
molecular dynamics simulations of mixtures of residues found in
protein cores.  The distribution (dashed line) was obtained from
more than $200$ jammed packings containing $N=24$ residues that were
generated by isotropically compressing the system to jamming onset.
The distribution of packing fractions from cores of proteins in the
Dunbrack database is shown by the solid line. The inset shows the
distribution of packing fractions for Ala (blue circles), Ile (green
crosses), Leu (red diamonds), Met (teal triangles), Phe (purple
solid line), and Val (black dotted line) separately from the packing
simulations.}
\label{phi_sim_aa}
\end{figure}

We also performed molecular dynamics simulations of residues confined
within a cubic box (with periodic boundary conditions) to determine
whether $\langle \phi \rangle_{\rm EH} \approx 0.55$ can be explained
by jamming of non-spherical objects~\cite{schreck12}. We studied
mixtures of $N$ residues with the number of Ala, Ile, Leu, Met, Phe,
and Val residues chosen from a weighted distribution that matched the
percentages found in protein cores.  (We focused focused on non-polar 
residues, but because Gly has no side chain and Cys
can form disulfide bonds, these were not included in our analyses.) We initialized the system to a
small packing fraction ($\phi_i = 10^{-3}$), set the bond lengths, bond
angles, backbone and side chain dihedral angles of each residue with
values from randomly chosen instances of the amino acid in the
Dunbrack database, and placed the residues in the simulation box with
random initial positions and orientations.

We then compressed the system while keeping the overlaps between
nonbonded atoms at approximately $10^{-6}$ by minimizing the enthalpy
$U+PV$ of the system, where $U$ is the total repulsive Lennard-Jones
potential energy between non-bonded atoms,
$P=10^{-6} \epsilon/\textrm{\AA}^{3}$ is the pressure of the
system, and $V$ is the volume of the simulation box. The algorithm
minimizes the enthalpy with respect to the variables $\vec{s}_{i}={\vec
  r}_{i}/V^{1/3}$ and logarithm of the box volume $\eta \propto \ln
(V/V_{0})$, where $V_0$ is the initial volume. Residue conformations
were strictly maintained using rigid body dynamics. We stopped the
minimization algorithm when the system was in force balance, with the
total force on each atom below the threshold value,
$\max_{i}\sum_{j}\left|\vec{F}_{ij}\right|<10^{-12}
\epsilon/\textrm{\AA}$ and final packing fraction $\phi_J$.

Fig.~\ref{phi_sim_aa} shows that the distribution of packing fractions $P(\phi_J)$
from the packing simulations is similar to the
distribution of packing fractions of protein cores from high
resolution protein crystal structures.  As an inset, we also show that
the packing fraction distribution for each residue from the
simulations is similar to that for the whole system.
 Fig.~\ref{phi_sim_aa} includes results for $N=24$ ($\sim 500$
atoms), but we found similar results for $N=8$ and $16$. These results
indicate that the connectivity of the protein backbone does not provide significant
constraints on the free volume in protein cores.

In summary, we have shown that using the explicit hydrogen hard-sphere
model for amino acids reproduces the side chain dihedral angle
distributions observed in protein crystal structures.  Moreover, we
find that the explicit hydrogen hard-sphere model gives a packing
fraction of $\langle \phi \rangle_{\rm EH} \approx 0.55$ for protein
cores. This value is similar to packing fractions for random packings
of non-spherical and elongated particles.  This result revises the
prior picture of protein cores as closely packed equal-sized spheres.
We believe that the revised packing fraction will serve as a target for
understanding the physical consequences of amino acid mutations and
the design of new proteins and interfaces.
 
We gratefully acknowledge the support of the Raymond and Beverly
Sackler Institute for Biological, Physical, and Engineering Sciences,
National Library of Medicine training grant T15LM00705628 (J.C.G.),
and National Science Foundation DMR-1307712 (L.R.), and also
benefited from the facilities and staff of the Yale University Faculty
of Arts and Sciences High Performance Computing Center and the NSF
(Grant No. CNS-0821132) that in part funded acquisition of the
computational facilities.

\bibliographystyle{unsrt}
\bibliography{ref_1}

\end{document}